\documentclass[10pt,twocolumn,journal]{IEEEtran}
\usepackage{amscd}
\usepackage{amssymb}
\usepackage{stmaryrd}
\usepackage{amsmath}
\usepackage{accents}
\usepackage{eepic, epic, pspicture}
\usepackage{array}

\usepackage{graphicx}
\usepackage{subfigure}
\usepackage{url}
\usepackage{color}
\usepackage{soul}
\usepackage{boxedminipage}
\usepackage[ruled,linesnumbered]{algorithm2e}
\usepackage{graphicx}
\usepackage{cite}

\newtheorem{theorem}{\textbf{Theorem}}
\newtheorem{lemma}{\textbf{Lemma}}
\newtheorem{problem}{\textbf{Problem}}

\begin{document}

\title{Energy Efficient Spectrum Sensing for State Estimation over A Wireless Channel}
\author{\IEEEauthorblockN{{Xianghui Cao\IEEEauthorrefmark{1},
Xiangwei Zhou\IEEEauthorrefmark{2} and Yu Cheng\IEEEauthorrefmark{1}}\\
\IEEEauthorblockA{\IEEEauthorrefmark{1}Department of ECE, Illinois Institute of Technology, Chicago, IL 60616.
Email: \{xcao10,cheng\}@iit.edu}\\
\IEEEauthorblockA{\IEEEauthorrefmark{2}Department of ECE, Southern Illinois University Carbondale, Carbondale, IL 62901.
Email: xzhou@engr.siu.edu}}}

\maketitle

\begin{abstract}
The performance of remote estimation over wireless channel is strongly affected by sensor data losses due to interference. Although the impact of interference can be alleviated by performing spectrum sensing and then transmitting only when the channel is clear, the introduction of spectrum sensing also incurs extra energy expenditure. In this paper, we investigate the problem of energy efficient spectrum sensing for state estimation of a general linear dynamic system, and formulate an optimization problem which minimizes the total sensor energy consumption while guaranteeing a desired level of estimation performance. The optimal solution is evaluated through both analytical and simulation results.
\end{abstract}

\begin{keywords}
Energy efficiency; Kalman filter; packet loss; spectrum sensing; state estimation
\end{keywords}

\section{Introduction}\label{introduction}
Estimating the state of dynamic processes is a fundamental task in many real-time applications such as environment monitoring, health-care, smart grid, industrial automation and wireless network operations \cite{Hespanhaetal2007,sukhavasi2013kalman}.
Consider remotely estimating the state of a general linear dynamic system, where sensor data are transmitted over a wireless channel to a remote estimator. Due to interference from other users on the same channel, the sensor data may randomly get lost, which can significantly affect the estimation performance \cite{Sinopoli_TAC:2004,KF_MarkovLoss:2007,rohr2011unified}.

To alleviate the impact of interference, a sensor can adopt the ``listen before talk" strategy, i.e., it can sense the channel first and only transmit data when the channel is clear. With spectrum sensing, the problem of estimation stability has been studied in \cite{ma2012state,JSAC:2014}, and the questions of whether and to what extent the state estimation performance can be improved have been addressed in \cite{JSAC:2014}. However, since both data transmission and spectrum sensing are energy consuming, the system energy efficiency becomes an important while challenging issue, which has not been studied in the literature yet.

In this paper, we investigate the problem of energy efficient spectrum sensing for state estimation over a wireless channel. Specifically, we consider when and how long to perform spectrum sensing in order to minimize the sensor's total energy consumption while guaranteeing a certain level of estimation performance. The problem is modeled as a mixed integer nonlinear programming (MINLP) which jointly optimizes the spectrum sensing frequency and sensing time, subjecting to an estimation performance constraint. The joint optimization in fact achieves a balance between spectrum sensing and transmission energy consumption. We derive a condition under which the estimation error covariance is stable in mean sense. Since the mean estimation error covariance is usually a random value and may vary slightly but not converge along time, we resort to a close approximation of the constraint which results in an approximated optimization problem whose solution suffices the original problem. Finally, we provide both analytical and simulation results of the solution to the optimization problem. The remainder of the paper is organized as follows. Section \ref{sec:overview} presents system model and optimization problem. The approximation problem is then introduced and analyzed in Section \ref{sec:singlechan}. Section \ref{sec:simulation} presents some simulation results, and Section \ref{sec:conclusion} concludes this paper.

\section{System Model and Problem Setup} \label{sec:overview}
We consider estimating the state of a general linear discrete-time dynamic process as follows.
\begin{align}\label{eq:state function}
\left\{
\begin{array}{lcl}
  x_{k+1} &=& A x_k + w_k,\\
  y_k &=& C x_k + v_k,
\end{array}\right.
\end{align}
where $x\in \mathbb{R}^{q_1}$ is the dynamic process state (e.g., environment variable) which changes along time. A wireless sensor is deployed to measure the process state and report the measurement to a remote estimator, where the sensor's measurement about $x$ is $y\in\mathbb{R}^{q_2}$. In the above, $q_1$ and $q_2$ are dimensions of $x$ and $y$, respectively. Note that the estimator only has noisy information of both process model and sensor measurements. The noises are denoted as $w_k$ and $v_k$ with $\mathbb{E}[w_k w^T_k]= Q$, $\mathbb{E}[v_k v^T_k]=R$ and $\mathbb{E}[w_i v^T_j]=0$, where $(\cdot)^T$ denotes the transpose of a matrix or vector. $A$ and $C$ are constant matrices. Assume that $C$ has full column rank and that $(A,Q^{\frac12})$ is controllable \cite{Sinopoli_TAC:2004}.

The sensor data are transmitted to a remote estimator where the transmissions are augmented by the spectrum sensing technique. The estimator applies a modified Kalman Filter \cite{Sinopoli_TAC:2004} to estimate the system state $x$ recursively. Given the system model as shown in (\ref{eq:state function}), define $\hat{x}_{k|k-1}$ and $\hat{x}_{k|k}$ as the prediction and estimate of the system state at step $k$, respectively. Define $P_{k|k-1} := \mathbb{E}[(x_k-\hat{x}_{k|k-1})(x_k-\hat{x}_{k|k-1})^T]$ and $P_{k|k} := \mathbb{E}[(x_k-\hat{x}_{k|k})(x_k-\hat{x}_{k|k})^T]$ as the covariance of the prediction and estimation errors, respectively. According to \cite{Sinopoli_TAC:2004}, the estimation process can be given as follows.
\begin{equation}
    \left\{
        \begin{array}{rcl}
            \hat{x}_{k|k-1} &=& A\hat{x}_{k-1|k-1}\\
            P_{k|k-1} &=& AP_{k-1|k-1}A^T + Q\\
            \hat{x}_{k|k} &=& \hat{x}_{k|k-1} + \gamma_k K_k(y_k-C\hat{x}_{k|k-1})\\
            K_k &:=& P_{k|k-1}C^T(CP_{k|k-1}C^T+R)^{-1}\\
            P_{k|k} &=& (I-\gamma_kK_k)P_{k|k-1}
        \end{array}
    \right.
\end{equation}
with a given initial value $P_{1|0} \geq 0$, where $I$ is an identity matrix of compatible dimension. In the above, $\gamma_k\in \{0,1\}$ represents whether the measurement packet is dropped or not in step $k$, i.e., $\gamma_k=1$ if successfully received and $\gamma_k=0$ otherwise. $\mathbb{P}[\gamma_k=0]$ characterizes the \emph{packet loss rate}.

Let $t_I$ and $t_B$ represent the idle and busy periods of the channel, respectively. We assume that \cite{IanF_TWC:2008}
\begin{align*}
    \Gamma_{I}(t) &= 1-e^{-\alpha t}\textrm{~and~} \Gamma_{B}(t) = 1-e^{-\beta t}.
\end{align*}
Thus, $\mathbb{E}[t_B]=\frac{1}{\beta}$, $\mathbb{E}[t_I]=\frac{1}{\alpha}$ and the idle and busy probabilities are $p_{I}=\frac{\beta}{\alpha+\beta}$ and $p_{B}=\frac{\alpha}{\alpha+\beta}$, respectively. Define $\eta$ as the probability that the channel will keep idle for at least $t_x$ period of time conditioned on that it is currently idle. We have
\begin{align*}
\nonumber &\eta=\frac{1}{p_{I}}\int^{0}_{-\infty} \mathbb{P}[\textrm{an idle period begins}]\left[1-\Gamma_{I}(t_x-t)\right]dt\\
\nonumber &= \frac{1}{p_{I}}\int^{\infty}_{0} \frac{1}{\frac1\alpha+\frac1\beta}\left[1-\Gamma_{I}(t_x+t)\right] dt\\
\nonumber &= \alpha\int^{\infty}_{0} e^{-\alpha (t_x+t)} dt= e^{-\alpha t_x}.
\end{align*}

We assume that the sensing time $\tau$ is bounded within $[0,\bar{\tau}]$ and is much smaller than both $\mathbb{E}[t_B]$ and $\mathbb{E}[t_I]$. Therefore, the channel state does not change during spectrum sensing (almost surely), and henceforth we can treat the sensing period as a point in time. The sampling period $T_s \gg \max(\mathbb{E}[t_B], \mathbb{E}[t_I])$, so that the packet drop rate in the current sampling period is irrelevant with that in previous steps. Based on this, the measurement packet drop rate, i.e., $\mathbb{P}[\gamma_k=0]$, also can be deemed time-independent.

Before transmitting a packet, the sensor must check the channel state and transmit packet only when the channel is available (in idle state). We adopt the \emph{energy detection} \cite{IanF_TWC:2008} as our spectrum sensing method. Let $s_c$ be the sensing outcome and define following two probabilities\footnote{In energy detection, whether the channel is idle is judged based on whether the detected energy is below a threshold $E_{th}$, referring to \cite{IanF_TWC:2008} for more details. Here, for simplicity, when the channel is idle, we assume $E_{th}=\epsilon_d \tau W \sigma^2_n$ where $\sigma_n$ is the channel noise power; otherwise, we assume $E_{th}=\epsilon_f \tau W (\sigma^2_s+\sigma^2_n)$ with $\sigma^2_s$ as the received signal power.}.
\begin{align}
\nonumber  p_d =\; &\mathbb{P}[s_c=\textrm{`idle'} | \textrm{channel is idle}]\\
         =\;& Q\left((1-\epsilon_d)\sqrt{\tau W}\right), \label{eq:pd}\\
\nonumber  p_f =\; &\mathbb{P}[s_c=\textrm{`idle'} | \textrm{channel is busy}]\\
         =\;& Q\left((1-{\epsilon_f})\sqrt{\tau W}\right), \label{eq:pf}
\end{align}
where $\epsilon_d>\epsilon_f>0$, $W$ is the channel bandwidth, and $Q(z) := \frac{1}{\sqrt{2\pi}} \int^{\infty}_{z} e^{-\frac{\tau^2}{2}} d\tau$. In the following, $p_d$ and $p_f$ are called the correct and false detection probabilities, respectively.

After sensing, the sensor will transmit packet only if the sensing result indicates an idle channel (we call this event a \emph{successful sensing}). Thus, the transmission probability is
\begin{align}\label{eq:ptx}
     p_{tx} &= p_{I}p_d + p_{B}p_f = \frac{1}{\alpha+\beta}(\beta p_d + \alpha p_f).
\end{align}

Define a sequence of variables $\{\theta_k\in \{0,1\}\}_{k\geq 1}$ as
\begin{equation}
    \theta_k = \left\{
        \begin{array}{rl}
            1, & \textrm{sense the channel in step $k$,}\\
            0, & \textrm{otherwise}.
        \end{array}
    \right.
\end{equation}

Let $\Theta := \{k|\theta_k = 1\}$, which is called the {spectrum sensing schedule}. In this paper, we restrict our attention to strict periodical spectrum sensing, i.e., $\Theta = \Theta_n :=\{0,n,2n,\ldots\}= \{k_i | k_i = in, i\in \mathbb{N}^+\cup\{0\}\}$, where $n$ represents the reciprocal of the sensing frequency.

\subsection{Problem Formulation}
Let $e_{s}$ and $e_{tx}$ denote the amounts of energy consumed by the sensor for conducting spectrum sensing in a unit time and transmitting a measurement packet (assume all packets are of the same length), respectively. If $\theta_k=1$, the average amount of energy consumed by the sensor in $k$th step is $\varphi_s  = \tau e_{s} + p_{tx}e_{tx}$. Therefore, under schedule $\Theta_n$, the average energy consumption in a single step is
\begin{align}\label{eq:energymodel}
    \bar{\varphi} = \frac{1}{n} \varphi_s= \frac{1}{n}(\tau e_{s} + p_{tx}e_{tx}).
\end{align}

The estimation performance can be characterizes by the error covariance $P_{k|k-1}$. For ease of exposition, hereafter, we let $P_k:=P_{k|k-1}$. Based on the estimation process above, we can see that $P_k$ is a function of the random variable $\gamma_k$; hence it is both random and time-varying and may not converge along an infinite horizon. Therefore, we consider the long-time average of the expected $P_k$, i.e., $\frac{1}{L}\sum^{L}_{k=1}\mathbb{E}[P_k]$, where $L$ is a sufficiently large number. We aim to bound this average value below a user defined threshold $\bar{P}$. With this constraint, our optimization problem can be formulated as follows.
\begin{problem}\label{problem:optimization}
Find the optimal schedule $\Theta_n$ and spectrum sensing time $\tau$ to
\begin{equation}\label{eq:problem2}
\left\{
\begin{array}{rl}
        \min\limits_{n,\tau}  &\; \bar{\varphi}= \frac{1}{n}(\tau e_{s} + p_{tx}e_{tx}) \\
        s.t.  &\; \frac{1}{L}\sum^{L}_{k=1}\mathbb{E}[P_k]\leq \bar{P}\\
              &\; 0\leq \tau  \leq \bar{\tau}.
\end{array}\right.
\end{equation}
\end{problem}

As can be seen, Problem \ref{problem:optimization} is a mixed integer nonlinear programming. Note that, through the joint optimization, the sensing energy and transmission energy are balanced.

\section{Main Results}\label{sec:singlechan}

\subsection{Estimation Stability}
To satisfy the constraints in (\ref{eq:problem2}), the sequence $\{\mathbb{E}[P_k]\}$ must be stable, i.e., $\mathbb{E}[P_k]<\infty, \forall k\geq 1$. For any $k\geq 1$, if $\theta_k=1$, based on the estimation process above, we have
\begin{align}\label{eq:PkPkm1}
\nonumber    P_k=\;&AP_{k-1}A^T+Q \\
\nonumber       & -\gamma_kAP_{k-1}C^T(CP_{k-1}C^T+R)^{-1}CP_{k-1}A^T \\
\nonumber      =\;& (1-\gamma_k)AP_{k-1}A^T+Q\\
\nonumber       &+\gamma_kA(P^{-1}_{k-1}+C^TR^{-1}C)^{-1}A^T\\
                =\;& (1-\gamma_k)AP_{k-1}A^T+Q + \gamma_kA\Upsilon_{k-1}A^T,
\end{align}
where $\Upsilon_{k-1}=(P^{-1}_{k-1}+C^TR^{-1}C)^{-1}$ is upper-bounded by $(C^TR^{-1}C)^{-1}$ (notice that $C$ has full column rank) \cite{JSAC:2014}.

Otherwise, $\theta_k=0$, which is similar to the case that the measurement packet gets lost. Then, $P_k=AP_{k-1}A^T+Q$. Consider the schedule $\Theta_n$. We have
\begin{align}
\nonumber    P_{k_i-1}=\;& AP_{k_i-2}A^T+Q = \ldots \\
                =\;& A^{n-1}P_{k_{i-1}}(A^T)^{n-1}+\sum^{n-2}_{t=0}A^tQ(A^T)^t.
\end{align}
Substituting the above equation into (\ref{eq:PkPkm1}) yields
\begin{align}
\nonumber    P_{k_i}=\;&(1-\gamma_{k_i})A^{n}P_{k_{i-1}}(A^T)^{n}+(1-\gamma_{k_i})\sum^{n-1}_{t=1}A^tQ(A^T)^t\\
                    &+Q + \gamma_{k_i}A\Upsilon_{k_i-1}A^T, \label{eq:Pki}\\
\nonumber    \mathbb{E}[P_{k_i}]=\;&(1-\gamma)A^{n}\mathbb{E}[P_{k_{i-1}}](A^T)^{n}+(1-\gamma)\sum^{n-1}_{t=1}A^tQ(A^T)^t\\
                        & +Q + \gamma A\mathbb{E}[\Upsilon_{k_i-1}]A^T, \label{eq:EPki}
\end{align}
where $\gamma$ is the \emph{successful packet reception rate} under $\theta_k=1$, which can be calculated by
\begin{align}\label{eq:gamma}
\nonumber    \gamma &=\mathbb{P}[\gamma_k=1|\theta_k=1,s_{c,k}=\textrm{`idle'}]\\
            &=p_I\eta p_d = \frac{\beta}{\alpha+\beta} p_d e^{-\alpha t_x},
\end{align}
where $s_{c,k}$ is the spectrum sensing result. Since $n$ is a finite constant, the stability of $\{\mathbb{E}[P_k]\}$ is equivalent to that of the original sequence $\{\mathbb{E}[P_{k_i}]\}$. Moreover, since $\Upsilon_{k_i-1}$ is bounded by a constant, the stability of $\{\mathbb{E}[P_{k_i}]\}$ is further equivalent to that of $\{X_{k_i}|X_{k_i}=(1-\gamma)A^{n}X_{k_{i-1}}(A^T)^{n}+\sum^{n-1}_{t=0}A^tQ(A^T)^t\}$. Therefore, it is easy to obtain the following condition which is both necessary and sufficient for the stability of $\{\mathbb{E}[P_k]\}$.

\begin{theorem}\label{theorem:convergibility n}
    $\forall\ n\geq 1$, $\{\mathbb{E}[P_k]\}$ is stable if and only if
    \begin{align}\label{eq:convergibility}
          (1-\gamma)\lambda_{\max}^{2n}(A) < 1,
    \end{align}
    where $\lambda_{\max}(\cdot)$ is the maximum eigenvalue of a square matrix.
\end{theorem}

Since $p_d\leq 1$, (\ref{eq:gamma}) indicates that $\gamma\leq \frac{\beta}{\alpha+\beta}e^{-\alpha t_x}<\frac{\beta}{\alpha+\beta}$. Therefore, an upper bound of $n$ can be obtained based on (\ref{eq:convergibility}) as follows.
\begin{equation}\label{eq:nbound}
    n\leq\bar{n}_1=\left\{
    \begin{array}{ll}
        \lceil\frac{\ln(\alpha+\beta)-\ln\alpha}{2\ln(\lambda_{\max}(A))}\rceil-1, & \textrm{ if } \lambda_{\max}(A)>1 \\
        \infty, & \textrm{ otherwise}.
    \end{array}\right.
\end{equation}

\subsection{Problem Approximation}\label{sec:approximation}
As shown in (\ref{eq:PkPkm1}), since $P_{k-1}$ appears in the inverse term of $\Upsilon_{k-1}$, $\mathbb{E}[P_k]$ will depend on all possible values of the sequence $\{\gamma_k\}_{k\geq 1}$. Moreover, $\mathbb{E}[P_k]$ may not necessarily converge. As a result, it is mathematically difficult to obtain the long-term average of $\mathbb{E}[P_k]$. Therefore, we resort to an upper bound of $\mathbb{E}[P_k]$ to sufficiently satisfy the constraint in Problem \ref{problem:optimization}. Based on Theorem 4 in \cite{Sinopoli_TAC:2004}, we have
\begin{align}
\nonumber    \mathbb{E}[P_k]=\;&\mathbb{E}[-\gamma_kAP_{k-1}C^T(CP_{k-1}C^T+R)^{-1}CP_{k-1}A^T]\\
\nonumber           &+\mathbb{E}[AP_{k-1}A^T+Q]\\
\nonumber          \leq &-\gamma A\mathbb{E}[P_{k-1}]C^T(C\mathbb{E}[P_{k-1}]C^T+R)^{-1}C\mathbb{E}[P_{k-1}]A^T\\
                        &+ A\mathbb{E}[P_{k-1}]A^T+Q.
\end{align}

Define a sequence $\{Y_k\}$ with
\begin{align}
\nonumber   Y_k =\;& AY_{k-1}A^T+Q \\
        &-\theta_k\gamma AY_{k-1}C^T(CY_{k-1}C^T+R)^{-1}CY_{k-1}A^T.
\end{align}
Then, $\mathbb{E}[P_k]\leq Y_k$ if we let $Y_0=P_0$. Lemma \ref{lemma:barY} characterizes the sequence $\{Y_k\}$; its proof is omitted due to limited space.
\begin{lemma}\label{lemma:barY}
    If (\ref{eq:convergibility}) holds, $\exists \bar{Y}(\gamma,n)>0$ such that
    \begin{equation}
        \lim_{L\to \infty}\frac{1}{L}\sum^L_{k=1}Y_k=\bar{Y}(\gamma,n).
    \end{equation}
    $\bar{Y}(\gamma,n)$ is monotonically decreasing as either $\gamma$ increases or $n$ decreases. Furthermore, for a sufficiently large $L$,
    \begin{align}
       \frac{1}{L}\sum^{L}_{k=1}\mathbb{E}[P_k] \leq \frac{1}{L}\sum^L_{k=1}Y_k \rightarrow \bar{Y}.
    \end{align}
\end{lemma}

Based on Lemma \ref{lemma:barY}, the constraint in Problem \ref{problem:optimization} can be approximated as $\bar{Y}(\gamma,n)\leq \bar{P}$. Due to the monotonicity of $\bar{Y}(\gamma,n)$ in $\gamma$, it is equivalent to say that $\gamma\geq \underline{\gamma}(n)$ where $\underline{\gamma}(n)$ is the unique solution of $\gamma$ to $\bar{Y}(\gamma,n)= \bar{P}$. On the other hand, since $\gamma\leq \frac{\beta}{\alpha+\beta}e^{-\alpha t_x}$, the inequality $\bar{Y}(\frac{\beta}{\alpha+\beta}e^{-\alpha t_x},n)\leq \bar{Y}(\gamma,n)\leq \bar{P}$ yields another upper bound on $n$:
\begin{equation}\label{eq:barn2}
    n\leq \bar{n}_2=\max\left\{\tilde{n}\bigl|\bar{Y}\left(\frac{\beta}{\alpha+\beta}e^{-\alpha t_x},\tilde{n}\right)\leq \bar{P}\right\}<\infty.
\end{equation}

Therefore, we get an approximation of Problem \ref{problem:optimization} as below.
\begin{problem}\label{problem:optimization_approx}
Find the optimal schedule $\Theta_n$ and spectrum sensing time $\tau$ to
\begin{align}
\left\{
\begin{array}{rl}
        \min\limits_{\Theta_n,\tau}\quad  &\; \bar{\varphi}= \frac{1}{n}(\tau e_{s} + p_{tx}e_{tx})\\
        s.t.\quad  &\; \gamma\geq \underline{\gamma}(n)\\
              &\; n\leq \bar{n}=\min\{\bar{n}_1,\bar{n}_2\}\\
              &\; 0\leq \tau  \leq \bar{\tau}.
\end{array}\right.
\end{align}
\end{problem}

\subsection{Optimal Solution Analysis}\label{sec:optimalsol1chan}
Given any $n$, Problem \ref{problem:optimization_approx} reduces to a subproblem with $\tau$ as the only decision variable. Since $n<\bar{n}$, the optimal $n^*$ and $\tau^*$ can be obtained by solving $\bar{n}$ such subproblems. In the following, we analyze the optimal solution $\tau^*_n$ under any given $n$. Let $\rho=\frac{\alpha}{\beta}$. We focus on that $\rho<1$, while the case that $\rho\geq 1$ can be analyzed in the same way. For ease of analysis, we assume $\tau$ is continuous. Given $n$, the subproblem has following properties.
\begin{align}
        \frac{\partial \gamma}{\partial \tau} 
        =\;& \frac{(1-p_1)\sqrt{W}}{2\sqrt{2\pi\tau}}(\epsilon_d-1)e^{-\frac{(1-\epsilon_d)^2}{2}W\tau}\label{eq:partialgamma}\\
        \frac{\partial \bar{\varphi}}{\partial \tau} 
     =\;& \frac{1}{n}\left(e_{s} + \frac{e_{tx}\sqrt{W}}{2(1+\rho)\sqrt{2\pi\tau}}f(\epsilon_d,\epsilon_f,\tau)\right)
            \label{eq:partialvarphi}\\
\nonumber     f(\epsilon_d,\epsilon_f,\tau) \triangleq& (\epsilon_d-1)e^{-\frac{(1-\epsilon_d)^2}{2}W\tau}\\
      & - \rho (1-\epsilon_f)e^{-\frac{(1-\epsilon_f)^2}{2}W\tau}.\label{eq:mathcalL}
\end{align}

Depending on the values of $\epsilon_d$ and $\epsilon_f$ (note that $\epsilon_f < \epsilon_d$), the shapes of the $\gamma$ and $\bar{\varphi}$ curves are described as follows.

\begin{figure}[!ht]
\vspace{-1mm}
\centering
        \subfigure[Case 1]{
            \includegraphics*[scale=0.26, bb= 106 221 505 603]{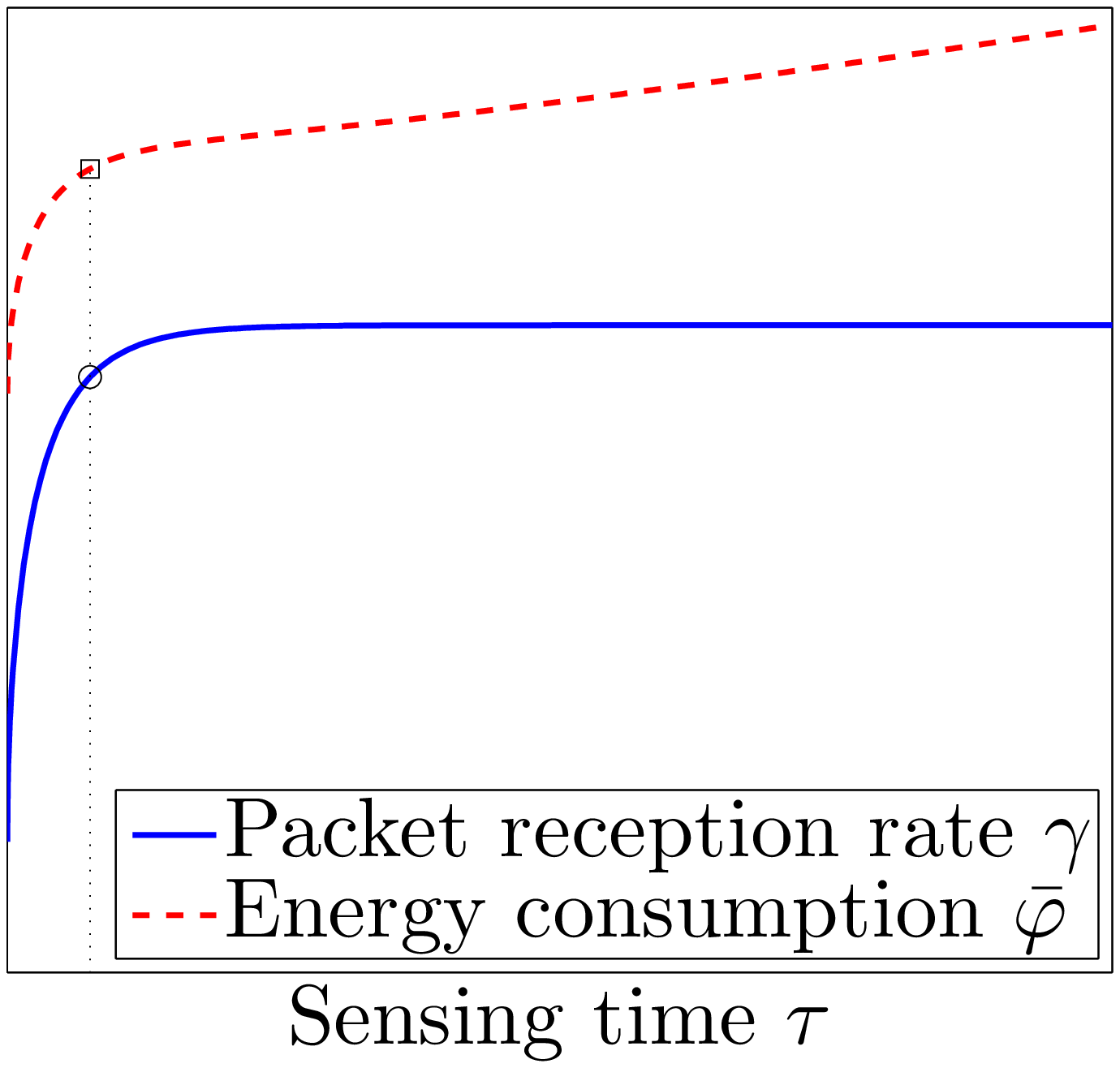}
            \label{Fig:epsilon1}
        }
        \hspace{-1mm}
        \subfigure[Case 2]{
            \includegraphics*[scale=0.26, bb= 106 221 505 603]{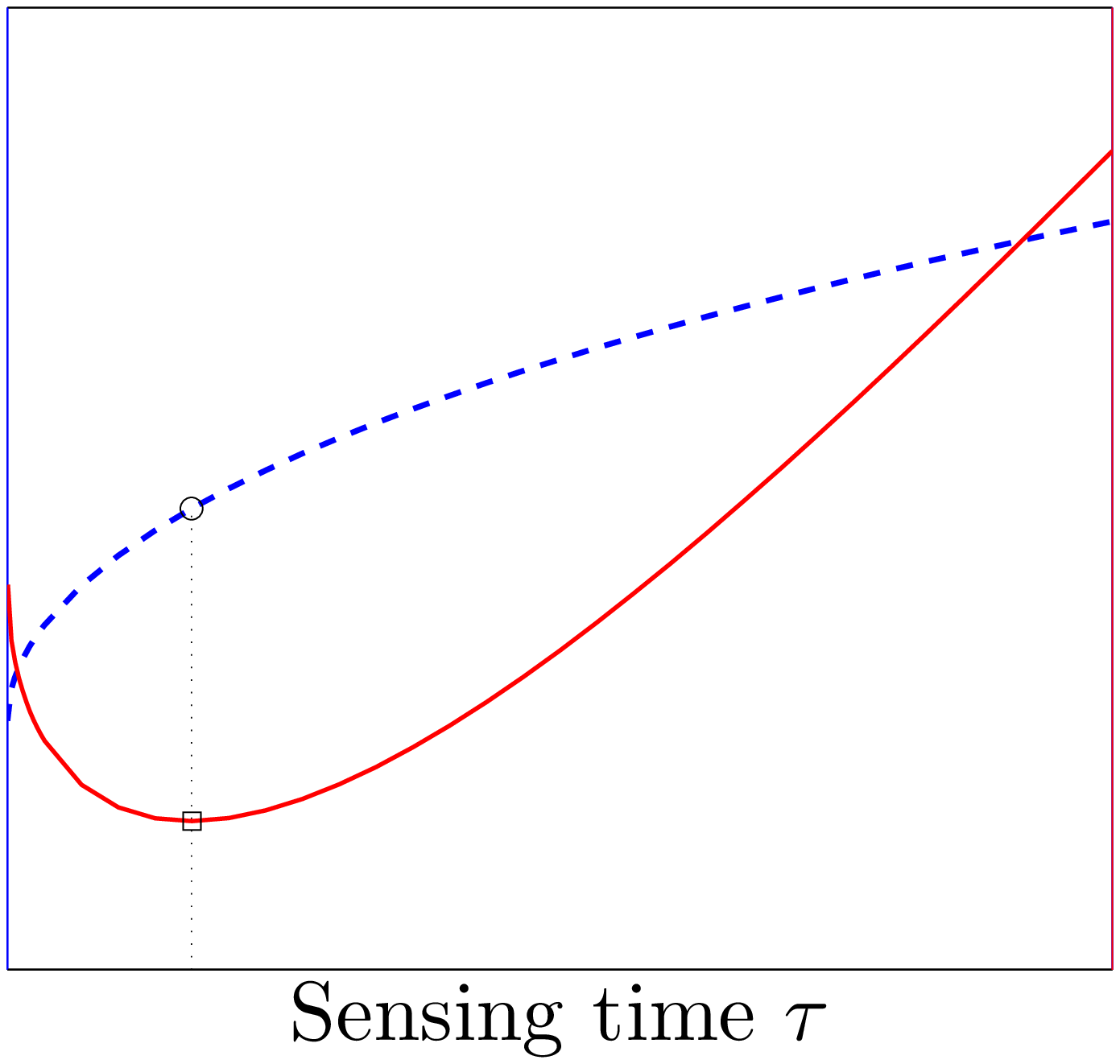}
            \label{Fig:epsilon2}
        }
        \subfigure[Case 3]{
            \includegraphics*[scale=0.26, bb= 106 221 505 603]{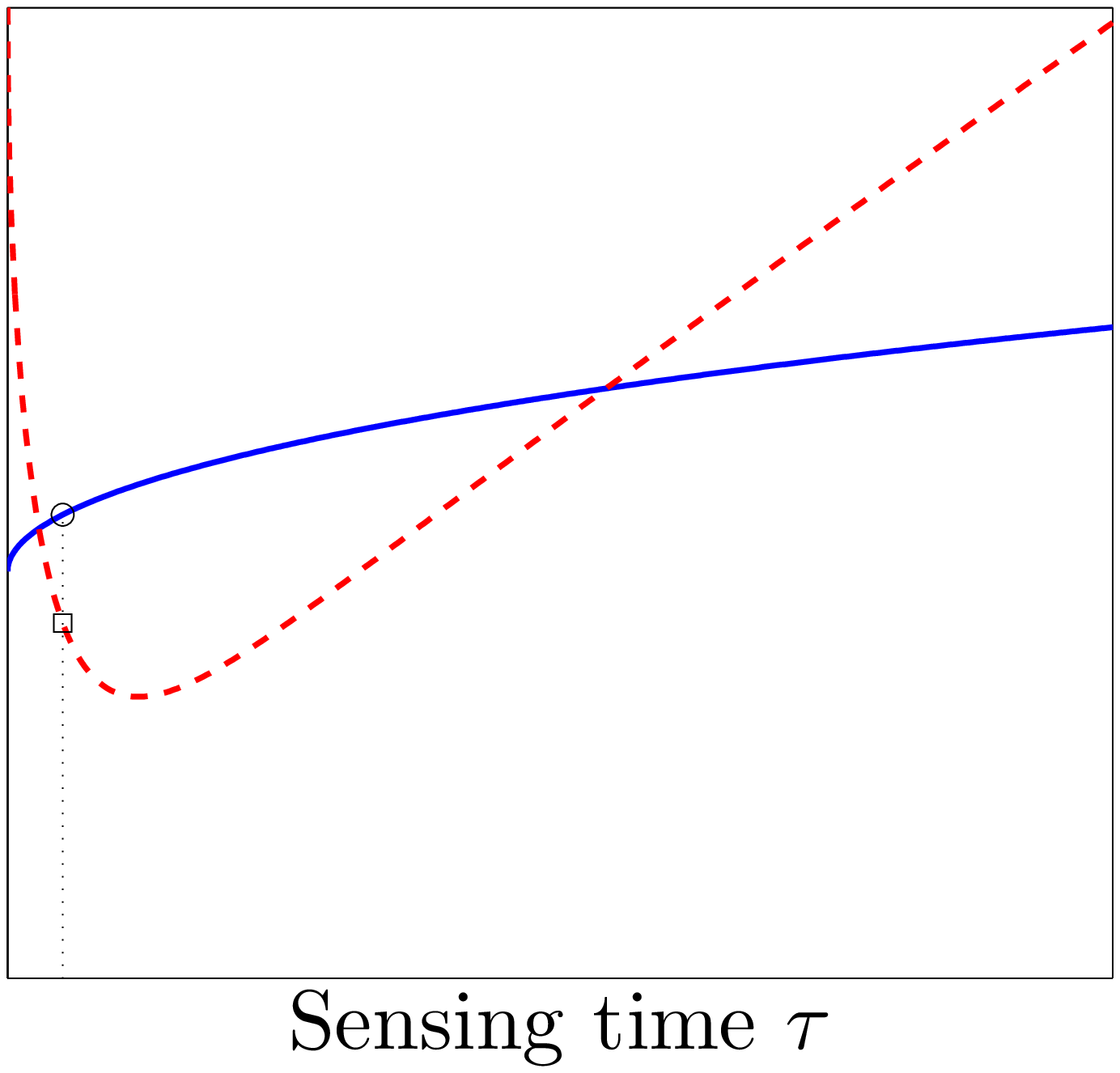}
            \label{Fig:epsilon3}
        }
        \hspace{-1mm}
        \subfigure[Case 4]{
            \includegraphics*[scale=0.26, bb= 106 221 505 603]{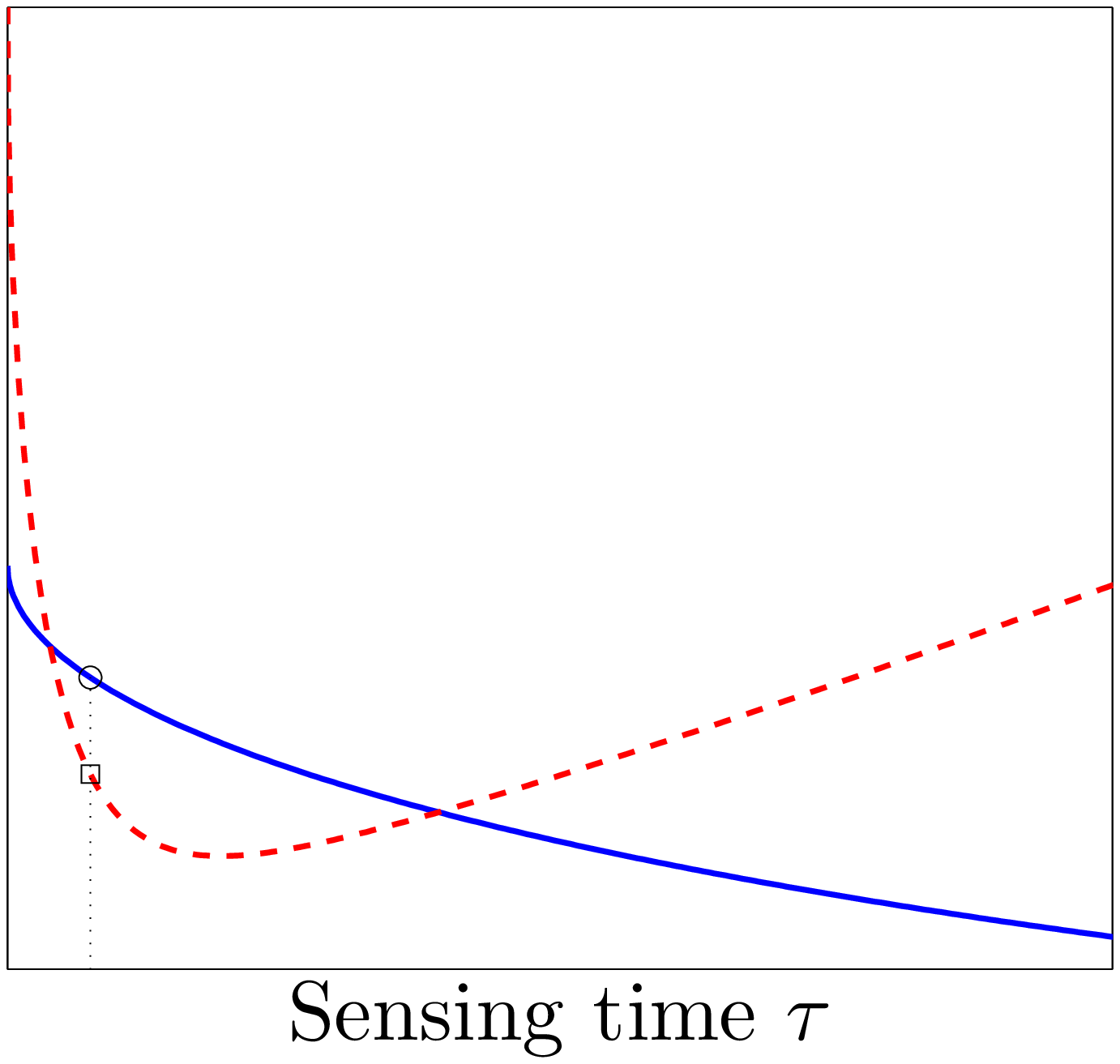}
            \label{Fig:epsilon4}
        }
\vspace{-2mm}
\caption{Illustrations of the optimal $\tau$ under different $\epsilon_d$ and $\epsilon_f$.} \label{Fig:epsilon}
\end{figure}

1) If either $\epsilon_d\geq 1$ and $\epsilon_f\geq 1$ or $\rho\leq\frac{\epsilon_d-1}{1-\epsilon_f}\leq 1$ and $\epsilon_f< 1$, it is easy to see that $\frac{\partial \gamma}{\partial \tau}\geq0$ and $\frac{\partial \bar{\varphi}}{\partial \tau}\geq0$, which means that both $\gamma$ and $\bar{\varphi}$ are increasing as $\tau$ increases. This corresponds to case 1 as shown in Fig. \ref{Fig:epsilon1}.

2) If $\frac{\epsilon_d-1}{1-\epsilon_f}> 1$ and $\epsilon_f< 1$, since $e^{-\frac{(1-\epsilon_d)^2}{2}W\tau}<e^{-\frac{(1-\epsilon_f)^2}{2}W\tau}$, $f(\epsilon_d,\epsilon_f,\tau)$ varies from positive infinite to a negative value and finally converges to 0. Depending on the parameters such as $e_{s}$ and $e_{tx}$, the shape of $\bar{\varphi}$ will be in the form of either case 1 or case 2 as shown in Fig. \ref{Fig:epsilon2}.

3) If $0\leq\epsilon_d-1<\rho(1-\epsilon_f)$, one can verify that $\frac{\partial^2}{\partial^2 \tau}f(\epsilon_d,\epsilon_f,\tau)>0$; hence, $\frac{\partial \bar{\varphi}}{\partial \tau}$ increases from negative infinite to a positive value. Therefore, as shown in Fig. \ref{Fig:epsilon3}, $\bar{\varphi}$ is a convex function.

4) Otherwise, $\epsilon_d<1$. Then, $\epsilon_f<1$ either. Consequently, $\frac{\partial \gamma}{\partial \tau}<0$ and $\frac{\partial}{\partial \tau}f(\epsilon_d,\epsilon_f,\tau)<0$. As shown in Fig. \ref{Fig:epsilon4}, the objective function is convex.

As shown in the figure, in case 1, the optimal $\tau^*_n$ is the smaller one between $\bar{\tau}$ and the point where $\gamma = \underline{\gamma}(n)$. In the other cases, let $\tau_{n,\bar{\varphi}}$ and $\tau_{n,\gamma}$ be the solution points for $\frac{\partial \bar{\varphi}}{\partial \tau}=0$ and $\gamma = \underline{\gamma}(n)$, respectively. In case 2, $\tau^*_n$ is among $\{0,\tau_{n,\bar{\varphi}}, \tau_{n,\gamma},\bar{\tau}\}$. In the other cases, $\tau^*_n\in\{\tau_{n,\bar{\varphi}}, \tau_{n,\gamma},\bar{\tau}\}$.

\section{Simulation Results}\label{sec:simulation}
In our simulations, we consider a linear system (\ref{eq:state function}) with $A=\begin{bmatrix}1.05& 0\\ 1& 0.9\end{bmatrix}$, $C=I$, $Q = I$ and $R = 0.8I$, where $I$ is the 2-by-2 identity matrix. The sensor samples the system every $T_s=1$ second and the transmission time of each measurement packet is $t_x=50ms$. The wireless channel has bandwidth $W=2Mbps$, noise power $\sigma_n=1$ and signal-to-noise ratio $-3dB$. The default average busy and idle rates for the channel are $\alpha=5$ and $\beta=20$, respectively. Other parameters are: $\epsilon_d=1.2$, $\bar{\tau}=20ms$, $e_s=e_{tx}=100$. The estimation performance requirement is set as $\bar{P}=\bar{Y}(0.7,6)$, where $\bar{Y}(\gamma,n)$ is defined in Lemma \ref{lemma:barY}.

\begin{figure}[!ht]
\vspace{-1mm}
\centering
        \subfigure[]{
              \includegraphics*[scale=0.24, bb=50 178 541 633]{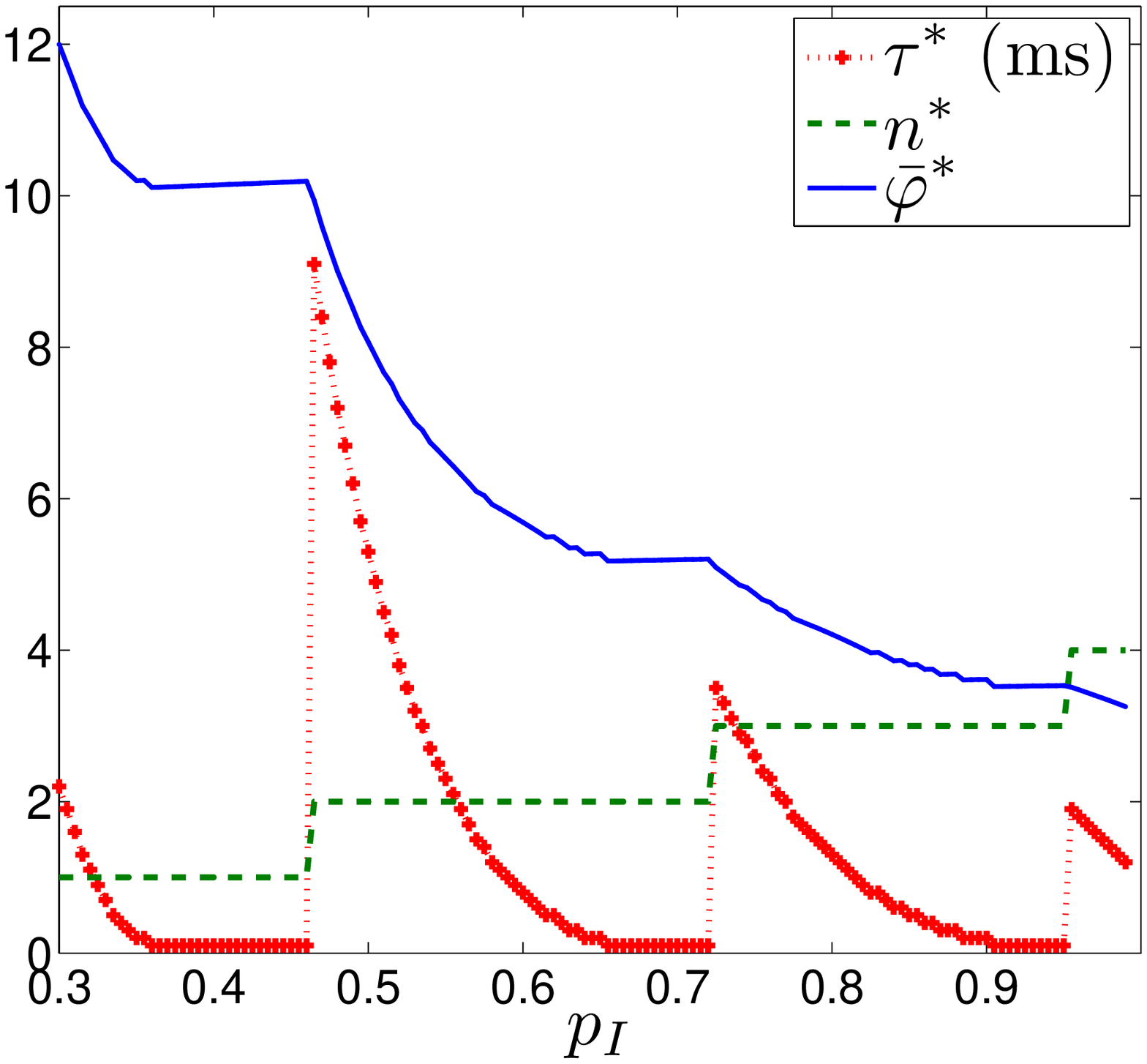}
            \label{Fig:PpI}
        }
        \hspace{-3mm}
        \subfigure[]{
            \includegraphics*[scale=0.24, bb=50 176 552 641]{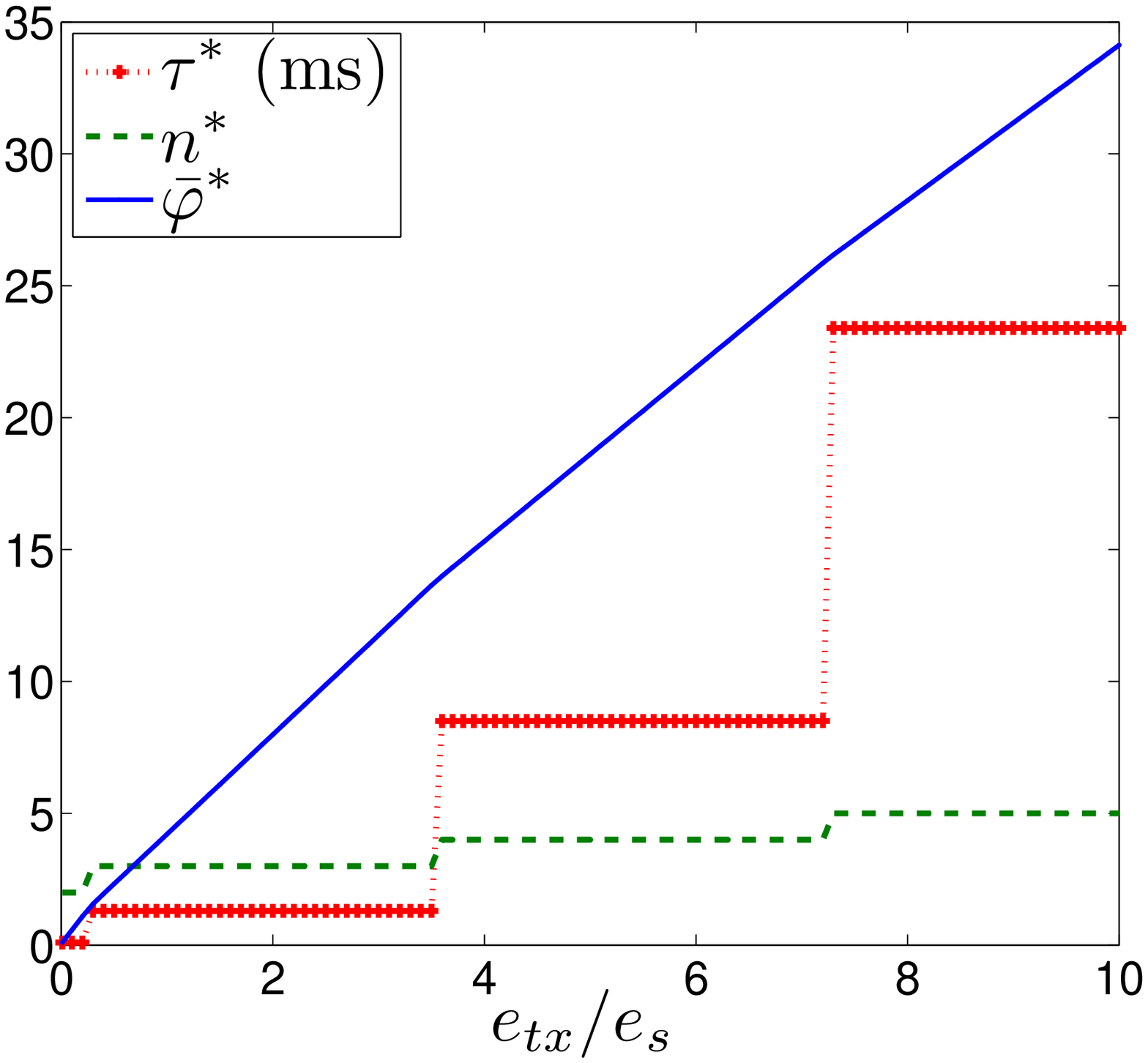}
            \label{Fig:Petx}
        }
\vspace{-2mm}
\caption{Optimal solutions under varying $p_I$ and $e_{tx}/e_s$.} \label{Fig:alg1}
\end{figure}

The optimal solutions of Problem \ref{problem:optimization_approx} are depicted in Fig. \ref{Fig:alg1}. In the left figure, we vary the channel idle probability $p_I$ by gradually increasing $\beta$. The results show that, under a certain $n$, the optimal sensing time $\tau^*$ drops quickly as the idle probability increases, which in turn results in the decrease of the average energy consumption $\bar{\varphi}$. In fact, as the channel quality becomes better, less sensor energy will be wasted for conduction unsuccessful sensing  and collided transmissions. Meanwhile, when $p_I$ increases from 0.3 to 1, the optimal $n$ increases piecewise, which means that the sensor conducts spectrum sensing and packet transmission less frequently. Therefore, generally speaking, the energy consumption decreases as $p_I$ increases.

The right figure demonstrate the optimal solutions under varying $e_{tx}/e_s$. As $e_{tx}$ increases, i.e., the transmission energy becomes to dominate the total energy $\bar{\varphi}$, the sensor's best strategy becomes to transmit data less frequently but more reliably in order to avoid collision and save energy. Therefore, it will use a larger $n$ and spend more sensing time to increase the sensing accuracy, which are clearly shown in Fig. \ref{Fig:Petx}.

\section{Conclusion}\label{sec:conclusion}
We have studied the energy efficient spectrum sensing problem for remote state estimation and formulated it as a mixed integer nonlinear programming problem. Both analytical and simulation results of the optimal solutions of the spectrum sensing time $\tau^*$ and period $n^*$ have been provided. We showed that, as $p_I$ increases, $n^*$ increases piecewise and the resulted energy consumption decreases. On the other hand, both $n^*$ and $\tau^*$ increase piecewise as $e_{tx}$ increases. Our future directions include extending the idea to multiple channel and multiple sensor scenarios.

\bibliographystyle{IEEEtran}
\bibliography{switch}

\end{document}